\documentclass[twocolumn,floatfix]{revtex4}
\usepackage{epsf,graphicx}
\usepackage{amsmath,amssymb,amsfonts}
\newcommand{\vecbm}[1]{\mbox{\boldmath#1}}

\newcommand{\nvec}[1]{\stackrel{\rightarrow}{#1}}

\begin{document}
\title{Microscopic statistical basis of classical thermodynamics of finite systems. }
\author{D.H.E. Gross}
\affiliation{Hahn-Meitner Institute and Freie Universit{\"a}t Berlin,
Fachbereich Physik. Glienickerstr. 100; 14109 Berlin, Germany}
\email{gross@hmi.de} \homepage{http://www.hmi.de/people/gross/ }
\begin{abstract}Heat can flow from cold to hot at any phase separation.
Therefore Lynden-Bell's gravo-thermal catastrophe
\cite{lyndenbell68} must be reconsidered. The original objects of
Thermodynamics, the separation of phases at first order phase
transitions, like boiling water in steam engines, are not
described by a single canonical ensemble. Inter-phase fluctuations
are not covered. The basic principles of statistical mechanics,
especially of phase transitions have to be reconsidered {\em
without the use of the thermodynamic limit}. Then
thermo-statistics applies also to nuclei and large astronomical
systems. A lot of similarity exists between the accessible phase
space of fragmenting nuclei and inhomogeneous multi stellar
systems.
\end{abstract}
\maketitle

Since the beginning of Thermodynamics in the first half of the
19.century its original motivation was the description of steam
engines and the liquid to gas transition of water. Here water
becomes inhomogeneous and develops a separation of the gas phase
from the liquid, i.e. water boils. This will be analyzed from the
new perspective of microcanonical statistics in section
\ref{chsplit}.

A little later statistical mechanics was developed by
Boltzmann\cite{boltzmann1872} to explain the microscopic
mechanical basis of Thermodynamics. Up to now it is generally
believed that this is given by the Boltzmann-Gibbs canonical
statistics. As traditional canonical statistics works only for
homogeneous, infinite systems, phase separations remain outside of
standard Boltzmann-Gibbs thermo-statistics, which, consequently,
signal phase-transitions of first order by Yang-Lee singularities.

It is amusing that this fact that is essential for the original purpose of
Thermodynamics to describe steam engines was never treated completely in
the past 150 years.  The system must be somewhat artificially split into
(still macroscopic and homogeneous) pieces of each individual phase
\cite{guggenheim67}. The most interesting configurations of two coexisting
phases cannot be described by a single canonical ensemble. Important
inter-phase fluctuations remain outside, etc. This is all hidden due to the
restriction to homogeneous systems in the thermodynamic limit.

Also the second law can rigorously be formulated only microcanonically:
Already Clausius \cite{clausius1854} distinguished between external and
internal entropy generating mechanisms. The second law is only related to
the latter mechanism \cite{prigogine71}, the internal entropy generation.
Again, canonical Boltzmann-Gibbs statistics is insensitive to this
important difference.

For this purpose, and also to describe small systems like fragmenting
nuclei or non-extensive ones like macroscopic systems at phase-separation,
or even very large, self-gravitating, systems, we need a new and deeper
definition of statistical mechanics and as the heart of it: of entropy. For
this purpose it is crucial to avoid the thermodynamic limit.
\section{What is entropy?} Entropy, S, is the characteristic
entity of thermodynamics. Its use distinguishes thermodynamics
from all other physics; therefore, its proper understanding is
essential. The understanding of entropy is sometimes obscured by
frequent use of the Boltzmann-Gibbs canonical ensemble, and the
thermodynamic limit. Also its relationship to the second law is
beset with confusion between external transfers of entropy $d_eS$
and its internal production $d_iS$.

The main source of the confusion is of course the lack of a clear
{\em microscopic and mechanical} understanding of the fundamental
quantities of thermodynamics like heat, external vs. internal
work, temperature, and last not least entropy, at the times of
Clausius and possibly even today.

Clausius \cite{clausius1854} defined  a quantity which he first called the
{\em ``value of metamorphosis'', in German `` Verwandlungswert'' } in
\cite{clausius1854}. Eleven years later he \cite{clausius1865} gave it the
name ``entropy'' $S$:
\begin{equation}
S_b-S_a=\int_a^b{\frac{dE}{T}},\label{entropy}
\end{equation} where $T$ is the absolute temperature of the body when the
momentary change is done, and $dE$ is the increment (positive resp.
negative) of all different forms of energy (heat and potential) put into
resp. taken out of the system.

From the observation that heat does not flow from cold to hot (see
section \ref{zerolaw}, formula \ref{02law}, however section
\ref{chsplit}) he went on to enunciate the second law as:
\begin{equation}
\Delta S=\oint{\frac{dE}{T}}\ge 0,\label{secondlaw}
\end{equation}
which Clausius called the "{\em uncompensated metamorphosis}". As will be
worked out in section \ref{chsplit} the second law as presented by
eq.(\ref{secondlaw}) remains valid even in cases where heat flows from low
to higher temperatures.

Prigogine \cite{prigogine71}, c.f. \cite{guggenheim67}, quite clearly
stated that the variation of $S$ with time is determined by two, crucially
different, mechanisms of its changes: the flow of entropy $d_eS$ to or from
the system under consideration; and its internal production $d_iS$. While
the first type of entropy change $d_eS$ (that effected by exchange of heat
$d_eQ$ with its surroundings) can be positive, negative or zero, the second
type of entropy change $d_iS$ is fundamentally related to its spontaneous
internal evolution (``Verwandlungen'', ``metamorphosis''
\cite{clausius1854}) of the system, and states the universal
irreversibility of spontaneous transitions. It can be only positive in any
spontaneous transformation.

Clausius gives an illuminating example in \cite{clausius1854}: When an
ideal gas suddenly streams under isolating conditions from a small vessel
with volume $V_1$ into a larger one ($V_2>V_1$), neither its internal
energy $U$, nor its temperature changes, nor external work done, but its
internal (Boltzmann-)entropy $S_i$ eq.(\ref{boltzmann0}) rises, by $\Delta
S=N\ln{(V_2/V_1)}$ . Only by compressing the gas (e.g. isentropically) and
creating heat $\Delta E=E_1[(V_2/V_1)^{2/3}-1]$ (which must be finally
drained) it can be brought back into its initial state. Then, however, the
entropy change in the cycle, as expressed by integral (\ref{secondlaw}), is
positive ($=N\ln{(V_2/V_1)}$). This is also a clear example for a
microcanonical situation where the entropy change by an irreversible
metamorphosis of the system is absolutely internal. It occurs  during the
first part of the cycle, the expansion, where there is no heat exchange
with the environment, and consequently no contribution to the
integral(\ref{secondlaw}). The construction by eq.(\ref{secondlaw}) is
correct though artificial. After completing the cycle the Boltzmann-entropy
of the gas is of course the same as initially. All this will become much
more clear by Boltzmann's microscopic definition of entropy, which will
moreover clarify its real {\em statistical} nature:

Boltzmann\cite{boltzmann1872} later defined the entropy of an
isolated system (for which the energy exchange with the
environment $d_eQ \equiv 0$) in terms of the sum of possible
configurations, $W$, which the system can assume consistent with
its constraints of given energy  and volume:\begin{equation}
\fbox{\fbox{\vecbm{S=k*lnW}}}\label{boltzmann0}
\end{equation}as written on Boltzmann's tomb-stone, with
\begin{equation}
W(E,N,V)= \int{\frac{d^{3N}\nvec{p}\;d^{3N}\nvec{q}}{N!(2\pi\hbar)^{3N}}
\epsilon_0\;\delta(E-H\{\nvec{q},\nvec{p}\})}\label{boltzmann}
\end{equation} in semi-classical approximation. $E$ is the total energy, $N$ is
the number of particles and $V$ the volume. Or, more appropriate for a
finite quantum-mechanical system:
\begin{equation}W(E,N,V)
=\sum{\scriptsize\begin{array}{ll}\mbox{all eigenstates n of H with given
N,$V$,}\\\mbox{and } E<E_n\le E+\epsilon_0
\end{array}}
\end{equation}
and $\epsilon_0\approx$ the macroscopic energy resolution.  This is still
up to day the deepest, most fundamental, and most simple definition of
entropy. {\em There is no need of the thermodynamic limit, no need of
concavity, extensivity and homogeneity.}  In its semi-classical
approximation, eq.(\ref{boltzmann}), $W(E,N,V,\cdots)$ simply measures the
area of the sub-manifold of points in the $6N$-dimensional phase-space
($\Gamma$-space) with prescribed energy $E$, particle number $N$, volume
$V$, and some other time invariant constraints which are here suppressed
for simplicity. Because it was Planck who coined it in this mathematical
form, I will call it the Boltzmann-Planck principle. It is further
important to notice that $S(E,N,V)$ is everywhere analytical in $E$
\cite{gross206}. In the microcanonical ensemble are no ``jumps'' or
multivaluedness in $S(E)$, independently of whether there are phase
transitions or not, in clear contrast to the canonical $S(T,N,V)$. A fact
which underlines the fundamental role of microcanonical statistics.

The Boltzmann-Planck formula has a simple but deep physical interpretation:
$W$ or $S$  measure our ignorance about the complete set of initial values
for all $6N$ microscopic degrees of freedom which are needed to specify the
$N$-body system unambiguously\cite{kilpatrick67}. To have complete
knowledge of the system we would need to know (within its semiclassical
approximation (\ref{boltzmann})) the initial positions and velocities of
all $N$ particles in the system, which means we would need to know a total
of $6N$ values. Then $W$ would be equal to one and the entropy, $S$, would
be zero. However, we usually only know the value of a few parameters that
change slowly with time, such as the energy, number of particles, volume
and so on. We generally know very little about the positions and velocities
of the particles. The manifold of all these points in the $6N$-dim. phase
space, consistent with the given macroscopic constraints of $E,N,V,\cdots$,
is the microcanonical ensemble, which has a well-defined geometrical size
$W$ and, by equation (\ref{boltzmann0}), a non-vanishing entropy,
$S(E,N,V,\cdots)$. The dependence of $S(E,N,V,\cdots)$ on its arguments
determines completely thermostatics and equilibrium thermodynamics.

Clearly, Hamiltonian (Liouvillean) dynamics of the system cannot
create the missing information about the initial values - i.e. the
entropy $S(E,N,V,\cdots)$ cannot decrease. As has been further
worked out in \cite{gross183,gross207} the inherent finite
resolution of the macroscopic description implies an increase of
$W$ or $S$ with time when an external constraint is relaxed. Such
is a statement of the second law of thermodynamics, which requires
that the {\em internal} production of entropy be positive for
every spontaneous process. Analysis of the consequences of the
second law by the microcanonical ensemble is appropriate because,
in an isolated system (which is the one relevant for the
microcanonical ensemble), the changes in total entropy must
represent the {\em internal} production of entropy, see above, and
there are no additional uncontrolled fluctuating energy exchanges
with the environment.

\section{The Zero'th Law in conventional extensive Thermodynamics \label{zerolaw}} In
conventional (extensive) thermodynamics thermal equilibrium of two
systems (1 \& 2) is established by bringing them into thermal
contact which allows free energy exchange. Equilibrium is
established when the total entropy
\begin{equation}
S_{1+2}(E,E_1)=S_1(E_1)+S_2(E-E_1)\label{eq1}
\end{equation}
is maximal. Under an energy flux $\Delta E_{2\to 1}$ from $2\to 1$ the
total entropy changes to lowest order in $\Delta E$ by
\begin{equation}
\Delta S_{1+2}|_E=(T_2-T_1)\Delta E_{2\to 1}.
\end{equation}
Consequently, a maximum of $S_{total}(E,E_1)|_E\ge S_{1+2}$ will be
approached when
\begin{equation}
 \mbox{sign}(\Delta S_{total})=\mbox{sign}(T_2-T_1)\mbox{sign}(\Delta E_{2\to
 1})>0.\label{02law}
\end{equation}
From here Clausius' first formulation of the Second Law follows:
"Heat always flows from hot to cold". Essential for this
conclusion is the {\em additivity} of $S$ under the split
(eq.\ref{eq1}). There are no correlations, which are destroyed
when an extensive system is split. Temperature is an appropriate
control parameter for extensive systems.

It is further easy to see that {\em the heat capacity of an extensive
system with $S(E,N)=2S(E/2,N/2)$ is necessarily positive}
\begin{equation}
C_V(E)=\partial E/\partial T=-~~\frac{(\partial S/\partial
E)^2}{\partial^2S/\partial E^2}>0:\label{negheat}
\end{equation}
The combination two pieces of $N/2$ particles each, but with
different energy per particle, one at $e_a=e_2-\Delta e/2$ and a
second at $e_b=e_2+\Delta e/2$, must lead to $S(E_2,N)\ge
S(E_a/2,N/2)+S(E_b/2,N/2)$, the simple algebraic sum of the
individual entropies because by combining the two pieces one
normally looses information. This, however, is equal to
$[S(E_a,N)+S(E_b,N)]/2$, thus $S(E_2,N)\ge[S(E_a,N)+S(E_b,N)]/2$.
I.e. {\em the entropy $S(E,N)$ of an extensive system is
necessarily concave, $\partial^2S/\partial E^2<0$} and eq.
\ref{negheat} follows. In the next section we will see that
therefore  {\em extensive systems cannot have phase separation,
the characteristic signal of transition of first order.}
\section{No phase separation without a convex, non-extensive
$S(E)$\label{chsplit}}

At phase separation the weight $e^{S(E)-E/T}$ of the configurations with
energy E in the definition of the canonical partition sum
\begin{equation}
Z(T)=\int_0^\infty{e^{S(E)-E/T}dE}\label{canonicweight}
\end{equation} becomes {\em bimodal}, at the transition temperature it has
two peaks, the liquid and the gas configurations which are
separated in energy by the latent heat. Consequently $S(E)$ must
be convex ($\partial^2 S/\partial E^2>0$, like $y=x^2$) and the
weight in (\ref{canonicweight}) has a minimum at $E_{min}$ between
the two pure phases. Of course, the minimum can only be seen in
the microcanonical ensemble where the energy is controlled and its
fluctuations forbidden. Otherwise, the system would fluctuate
between the two pure phases by an, for macroscopic systems even
macroscopic, energy $\Delta E\sim E_{lat}$ of the order of the
latent heat. The heat capacity is
\begin{equation}
C_V(E_{min})=\partial E/\partial T=-~~\frac{(\partial S/\partial
E)^2}{\partial^2S/\partial E^2}<0.
\end{equation}
I.e. {\em the convexity of $S(E)$ and the negative heat capacity
are the generic and necessary signals of
phase-separation\cite{gross174}}. It is amusing that this fact,
which is essential for the original purpose of Thermodynamics to
describe steam engines, seems never been really recognized in the
past 150 years. However,  such macroscopic energy fluctuations and
the resulting negative specific heat are already early discussed
in high-energy physics by Carlitz \cite{carlitz72}.

The existence of the negative heat capacity at phase separation
has a surprising but fundamental consequence: Combining two equal
systems with negative heat capacity and different energy per
particle, they will relax with a flow of energy from the lower to
the higher temperature! This is consistent with the naive picture
of an {\em energy equilibration}. Thus {\em Clausius' "energy
flows always from hot to cold", i.e. the dominant control-role of
the temperature in thermo-statistics as emphasized by Hertz
\cite{hertz10a} is violated}. Of course this shows quite clearly
that {\em unlike to extensive thermodynamics the temperature is
not the appropriate control parameter in non-extensive situations
like e.g. at phase separations, nuclear fragmentation, or stellar
systems.\cite{gross212}}

By the same reason the well known paradox of Antonov in astro-physics due
to the occurrence of negative heat capacities must be reconsidered: By
using standard arguments from extensive thermodynamics  Lynden-Bell
\cite{lyndenbell68} claims that a system $a$ with negative heat capacity
$C_a<0$ in gravitational contact with another $b$ with positive heat
capacity $C_b>0$ will be unstable: If initially $T_a>T_b$ the hotter system
$a$ transfers energy to the colder $b$ and by this both become even hotter!
If $C_b>-C_a$, $T_a$ rises faster than $T_b$ and if the heat capacities
don't change, this will go for ever. This is Lynden-Bells gravo-thermal
catastrophe. This is wrong because just the opposite happens, the hotter
$a$ may even {\em absorb} energy from the colder $b$ and both systems come
to equilibrium at the same intermediate temperature c.f.
\cite{gross203,gross212}. Negative heat can only occur in the
microcanonical ensemble.

As phase separation exists also in the thermodynamic limit, by the same
arguments as above {\em the curvature of $S(E)$ remains convex,
$\partial^2S/(\partial E)^2>0$. Consequently, the negative heat capacity
should also be seen in ordinary macroscopic systems studied in chemistry!}

Searching for example in Guggenheims book \cite{guggenheim67} one finds
some cryptic notes in \S 3 that the heat capacity of steam at saturation is
negative. No notice that {\em this is the generic effect at any phase
separation!} Therefore let me recapitulate in the next section how chemists
treat phase separation of macroscopic systems and then point out why this
does not work in non-extensive systems like fragmenting nuclei, at phase
separation in normal macroscopic systems, or large astronomical systems.

\section{Macroscopic systems in Chemistry \label{chemistry}}

Systems studied in chemical thermodynamics consist of several {\em
homogeneous macroscopic} phases $\alpha_1,\alpha_2,\cdots$
cf.\cite{guggenheim67}. Their mutual equilibrium must be
explicitly constructed from outside.

Each of these phases are assumed to be homogeneous and macroscopic (in the
"thermodynamic limit" ($N_\alpha\to\infty|_{\rho_\alpha=const}$)). There is
no common canonical ensemble for the entire system of the coexisting
phases. Only the canonical ensemble of {\em each} phase separately becomes
equivalent in the limit to its microcanonical counterpart.

The canonical partition sum of {\em each} phase $\alpha$ is
defined as the Laplace transform of the underlying  microcanonical
sum of states $W(E)_\alpha=e^{S_\alpha(E)}$
\cite{gross147,gross158}
\begin{equation}
Z_\alpha(T_\alpha)= \int_0^\infty e^{S_\alpha(E)-E/T_\alpha} dE.
\end{equation}
The mean canonical energy is
\begin{equation}
 <E_\alpha(T_\alpha)>=T_\alpha^2\partial \ln Z_\alpha(T_\alpha)/\partial T_\alpha.
\end{equation}
In chemical situations proper the assumption of homogeneous
macroscopic individual phases is of course acceptable. In the
thermodynamic limit ($N_\alpha\to\infty|_{\rho_\alpha=const}$) of
a {\em homogeneous} phase $\alpha$, the canonical energy\\
$<\!\!E_\alpha(T_\alpha)\!\!>$ becomes identical to the
microcanonical energy $E_\alpha$ when the temperature is
determined by $ T_\alpha^{-1}=\partial
S_\alpha(E,V_\alpha)/\partial E_\alpha$. The relative width of the
canonical energy is
\begin{equation}
\Delta
E(T)_\alpha=\frac{\sqrt{<E_\alpha^2>_T-<E_\alpha>_T^2}}{<E_\alpha>_T}\propto
\frac{1}{\sqrt{N_\alpha}}.
\end{equation}
The heat capacity at constant volume is
\begin{eqnarray}
C_\alpha|_{V_\alpha}&=&\frac{<E_\alpha^2>_{T_\alpha}-<E_\alpha>_{T_\alpha}^2}{T_\alpha^2}\ge
0.\label{specheat}
\end{eqnarray}

Only in the thermodynamic limit ($N_\alpha\to\infty|_{\rho_\alpha=const}$)
does the relative energy uncertainty $\Delta E_\alpha\rightarrow 0$, and
the canonical and the microcanonical ensembles for each homogeneous phase
($\alpha$) become equivalent. This equivalence is the only justification of
the canonical ensemble controlled by intensive temperature $T$, or chemical
potential $\mu$, or pressure $P$. I do not know of any microscopic
foundation of the canonical ensemble and intensive control parameters apart
from the limit.

The positiveness of any canonical $C_V(T)$ or $C_P(T)$
(\ref{specheat}) is of course the reason why the inhomogeneous
system of several coexisting phases ($\alpha_1 \& \alpha_2$) with
an overall {\em negative} heat capacity cannot be described by a
{\em single common} canonical distribution
\cite{gross159,gross174}.

This new but fundamental interpretation of thermo-statistics was
introduced to the chemistry community in \cite{gross186,gross212}.

\section{New kind of phases well seen in hot nuclei
or multi-star systems.\label{nuclearfrag}}

The new lesson to be learned is that if one defines the phases by
individual peaks \footnote{Here I do not mean irregularities of
the order of $N^{-1/3}$ due to the discreteness of the quantum
level distributions} in $e^{S(E)-E/T}$ in (\ref{canonicweight}),
then there exist also {\em inhomogeneous phases} like in
fragmented nuclei or stellar systems. The general concept of
thermo-statistics becomes enormously widened.

Now, certainly neither the phase of the whole multi-fragmented
nucleus nor the individual fragments themselves can be considered
as macroscopic homogeneous phases in the sense of chemical
thermodynamics (ChTh). Consequently, (ChTh) cannot and should not
be applied to fragmenting nuclei and the microcanonical
description is ultimately demanded. This becomes explicitly clear
by the fact that the configurations of a multi-fragmented nucleus
have a {\em negative} heat capacity at constant volume $C_V$ and
also at constant pressure $C_P$ (if at all a pressure can be
associated to nuclear fragmentation \cite{gross174}). Meanwhile a
huge amount of experimental evidences of negative heat capacities
has accumulated: Nuclear fragmentation e.g. \cite{dAgostino00},
atomic clusters e.g. \cite{schmidt01}, astrophysics e.g.
\cite{thirring70}, conventional macroscopic systems at phase
separation e.g.\cite{guggenheim67}.

The existence of well defined peaks (i.e. phases as defined above)
in the event distribution of nuclear fragmentation data is
demonstrated very nicely in \cite{pichon03} from various points of
view. A lot more physics about the mechanism of phase transitions
can be learned from such studies.

\section{Outlook}

It is a deep and fascinating aspect of {\em nuclear} fragmentation: First,
in nuclear fragmentation we can measure the {\em whole statistical
distribution} of the ensemble event by event including eventual inter-phase
fluctuations. Not only their mean values are of physical interest.
Statistical mechanics can be explored from its first microscopic principles
in any detail well away from the thermodynamic limit. Initiated by our
theoretical studies of nuclear fragmentation we found  the very general
appearance of a {\em backbending} caloric curve $T(E)$ corresponding to a
negative heat capacity, e.g.:\cite{gross84} [an early review in
\cite{gross95}] similar effects were proposed in the melting of atomic
clusters \cite{wales90}. Years later its existence in nuclear fragmentation
was verified experimentally \cite{dAgostino00}. However, the necessary
convexity of the entropy $S(E)$ at {\em any} phase separation seems to be
little known in thermodynamics. {\em Clausius' version of the second law
``heat always flows from hot to cold'' is in general violated at any phase
separation even in macroscopic systems.} Nowadays, the non-equivalence of
the micro- and the canonical ensemble at phase-separations is discussed by
many authors, see e.g. several relevant papers in \cite{dauxiois02}.

In nuclear fragmentation there may be other conserved control
parameters besides the energy: E.g. in the recent paper by Lopez
et al. \cite{lopez05} the importance of the total angular momentum
of the excited nucleus was emphasized. In this paper a bimodality,
i.e. phase separation, in the mass-asymmetry of the fragments is
demonstrated controlled by the transferred spin and not by
excitation energy. This is an interesting, though still
theoretical, example of the rich facets of the fragmentation phase
transition in {\em finite} systems which goes beyond the
liquid-gas transition and {\em does not exist in macroscopic
chemistry}. Angular momentum is also a very crucial control
parameter in stellar systems, see below.

Second, and this may be more important: For the first time phase
transitions to non-homogeneous phases can be studied where these
phases are within themselves composed of several nuclei. This
situation is very much analogous to multi star systems like
rotating double stars during intermediate times, when nuclear
burning prevents their final implosion. The occurrence of negative
heat capacities is an old well known peculiarity of the
thermo-statistics of self-gravitating systems
\cite{lyndenbell68,thirring70}. Also these cannot be described by
a canonical ensemble. It was shown in \cite{gross207,gross203} how
the {\em microcanonical} phase space of these self-gravitating
systems has the realistic configurations which are observed. Of
course, the question whether these systems really fill uniformly
this phase space, i.e. whether they are interim equilibrized or
not is not proven by this observation though it is rather likely.


\begin{thebibliography}{10}

\bibitem{lyndenbell68}
D.~Lynden-Bell and R.~Wood.
\newblock The gravo-thermal catastrophe in isothermal spheres and the onset of
  red-giant structure for stellar systems.
\newblock {\em Mon. Not. R. astr. Soc.}, 138:495, 1968.

\bibitem{boltzmann1872}
L.~Boltzmann.
\newblock Weitere Studien \"uber das W\"armegleichgewicht unter Gas-Molek\"ulen.
\newblock {\em Sitzungsbericht der Akadamie der Wissenschaften, Wien},
  66:275--370, 1872.

\bibitem{guggenheim67}
E.A. Guggenheim.
\newblock {\em Thermodynamics, An Advanced Treatment for Chemists and
  Physicists}.
\newblock North-Holland Personal Library, Amsterdam, 1967.

\bibitem{clausius1854}
R.~Clausius.
\newblock \"Uber eine ver\"anderte Form des zweiten Hauptsatzes der
  mechanischen W\"armetheorie.
\newblock {\em Annalen der Physik und Chemie}, 93:481--506, 1854.

\bibitem{prigogine71}
P. Glansdorff and I. Prigogine.
\newblock {\em Thermodynamic Theory of Structure, Stability and Fluctuations}.
\newblock{John Wiley\& Sons, London.} 1971

\bibitem{clausius1865}
R.~Clausius.
\newblock \"Uber verschiedene f\"ur die Anwendung bequeme Formen der
  Hauptgleichungen der me\-cha\-nischen W\"armetheorie.
\newblock {\em Annalen der Physik und Chemie}, 125:353--400, 1865.

\bibitem{gross206}
D.H.E. Gross.
\newblock The microcanonical entropy is multiply differentiable no dinosaurs
  in microcanonical gravitation: No special "microcanonical phase transitions".
\newblock /cond-mat/0403582:3.

\bibitem{kilpatrick67}
J.E. Kilpatrick.
\newblock Classical thermostatistics.
\newblock In H.~Eyring, editor, {\em Statistical Mechanics}, number~II,
  chapter~1, pages 1--52. Academic Press, New York, 1967.

\bibitem{gross183}
D.H.E. Gross.
\newblock Ensemble probabilistic equilibrium and non-equilibrium thermodynamics
 without the thermodynamic limit.
\newblock In Andrei Khrennikov, editor, {\em Foundations of Probability and
 Physics}, number XIII in PQ-QP: Quantum Probability, White Noise Analysis,
 pages 131--146, Boston, October 2001. ACM, World Scientific.

\bibitem{gross207}
D.H.E. Gross.
\newblock A new thermodynamics from nuclei to stars.
\newblock {\em Entropy}, 6:158--179,cond-mat/0505450 (2004).

\bibitem{gross174}
D.H.E. Gross.
\newblock {\em Microcanonical thermodynamics: Phase transitions in ``Small''
  systems}, volume~66 of {\em Lecture Notes in Physics}.
\newblock World Scientific, Singapore, 2001.

\bibitem{carlitz72}
R.D. Carlitz.
\newblock Hadronic matter at high density.
\newblock {\em Phys.Rev.D}, 5:3231--3242, 1972.

\bibitem{hertz10a}
P.~Hertz.
\newblock \"Uber die mechanische Begr\"undung der Thermodynamik II.
\newblock {\em Ann. Phys. (Leipzig)}, 33:537, 1910.

\bibitem{gross212}
D.H.E. Gross and J.F. Kenney.
\newblock The microcanonical thermodynamics of finite systems: The microscopic
  origin of condensation and phase separations; and the conditions for heat
 flow from lower to higher temperatures.
\newblock {\em Journal of Chemical Physics},  cond--mat/0503604, (2005).

\bibitem{gross203}
D.H.E. Gross.
\newblock Classical equilibrium thermostatistics, "sancta sanctorum of
statistical mechanics", from nuclei to stars.
\newblock {\em Physica A}340/1-3:76,(2004), cond--mat/0311418.

\bibitem{gross147}
O.~Schapiro, D.H.E. Gross, and A.~Ecker.
\newblock Microcanonical monte carlo.
\newblock{\em First International
  Conference on Monte Carlo and Quasi-Monte Carlo Methods in Scientific
  Computing}, volume 106, pages 346--353, Las Vegas, Nevada, 1995.

\bibitem{gross158}
D.H.E. Gross and M.E. Madjet.
\newblock Microcanonical vs. canonical thermodynamics.
\newblock cond-mat/9611192.

\bibitem{gross159}
D.H.E. Gross and M.E. Madjet.
\newblock Cluster fragmentation, a laboratory for thermodynamics and
 phase-transitions in particular.
\newblock {\em Proceedings of
  ''Similarities and Differences between Atomic Nuclei and Clusters''}, pages
  203--214, Tsukuba, Japan 1997. AIP

\bibitem{gross186}
D.H.E. Gross.
\newblock Geometric foundation of thermo-statistics, phase transitions, second
  law of thermodynamics, but without thermodynamic limit.
{\em PCCP},4:863(2002), cond-mat/0201235.

\bibitem{dAgostino00}
M.~D'Agostino et al.
\newblock Negative heat capacity in the critical region of nuclear
 fragmentation: an experimental evidence of the liquid-gas phase transition.
\newblock {\em Phys.Lett.B}, 473:219--225, 2000.

\bibitem{schmidt01}
M.~Schmidt, R.~Kusche, T.~Hippler, J.~Donges, W.~Kornm\"uller,
B.~von
  Issendorff, and H.~Haberland.
\newblock Negative heat capacity for a cluster of 147 sodium stoms.
\newblock {\em Phys.Rev.Lett.}, 86:1191--1194, 2001.

\bibitem{thirring70}
W.~Thirring.
\newblock Systems with negative specific heat.
\newblock {\em Z. f. Phys.}, 235:339--352, 1970.

\bibitem{pichon03}
M.Pichon and the INDRA and ALADIN collaboration.
\newblock Bimodality in binary Au + Au collisions from 60 to 100 MeV/u.
\newblock {\em Proceedings XLI Winter Meeting, Bormio,2003}, p.149.

\bibitem{gross84}
D.H.E. Gross, Y.M. Zheng, and H.~Massmann.
\newblock New kind of phase transition in hot nuclei.
\newblock {\em Phys.~Lett.}, B~200:397--400, 1987.

\bibitem{gross95}
D.H.E. Gross.
\newblock Statistical decay of very hot nuclei, the production of large
  clusters.
\newblock {\em Rep.Progr.Phys.}, 53:605--658, 1990.

\bibitem{wales90}
D.J. Wales and S.~Berry.
\newblock Freezing, melting, spinodals, and clusters.
\newblock {\em J. Chem. Phys.}, 92:4473--4482, 1990.

\bibitem{dauxiois02}
In T.Dauxois, S.Ruffo, E.Arimondo, and M.Wilkens, editors, {\em
Dynamics and
  Thermodynamics of Systems with Long Range Interactions}, Lecture Notes in
  Physics, 602, Heidelberg, 2002. Springer.

\bibitem{lopez05}
O.~Lopez, D.~Lacroix, and E.Vient.
\newblock Bimodality as signal of liquid-gas phase transition in nuclei?
\newblock  nucl--th/0504027.

\end{thebibliography}

\end{document}